\documentclass[submission,copyright,creativecommons]{eptcs}

\providecommand{\event}{PLACES 2019}

\usepackage[utf8]{inputenc}
\usepackage[T1]{fontenc}
\usepackage{color}
\usepackage{listings}
\usepackage{tikz}
\usepackage{wrapfig}
\usepackage{amsmath,amssymb,stmaryrd}
\newcommand{\freest}{\textsc{FreeST}}

\newcommand\tcLolli\multimap
\newcommand\tcFun\to

\newcommand\Small{\small}

\definecolor{darkviolet}{rgb}{0.5,0,0.4}
\definecolor{darkgreen}{rgb}{0,0.4,0.2} 
\definecolor{darkblue}{rgb}{0.1,0.1,0.9}
\definecolor{darkgrey}{rgb}{0.5,0.5,0.5}
\definecolor{lightblue}{rgb}{0.4,0.4,1}

\lstdefinestyle{eclipse}{
  breaklines=true,
  basicstyle=\sffamily\Small,
  emphstyle=\color{red}\bfseries, 
  keywordstyle=\color{darkviolet}\bfseries,
  commentstyle=\color{darkgreen},
  stringstyle=\color{darkblue},
  numberstyle=\color{darkgrey},%\lstfontfamily,
  emphstyle=\color{red},
  morecomment=[s][\color{lightblue}]{/**}{*/},
  showstringspaces=false,
  numbers=left,
  tabsize=2
}

\lstdefinestyle{eclipse-Haskell}{
  breaklines=true,
  basicstyle=\sffamily\Small,
  emphstyle=\color{red}\bfseries, 
  keywordstyle=\color{darkviolet}\bfseries,
  commentstyle=\color{darkgreen},
  stringstyle=\color{darkblue},
  emphstyle=\color{red},
  morecomment=[s][\color{lightblue}]{/**}{*/},
  showstringspaces=false,
  numbers=none,
  tabsize=2
}

\lstdefinelanguage{freest}{
  style=eclipse,
  morekeywords=[1]{Int, Char, Bool, Skip, type, dualof, forall, rec, let, in, if, then, else, new, send, receive, select, fork, case, of, data, match, with, True, False},
  sensitive=true,
  literate=
  {->}{$\rightarrow$}2
  {-o}{$\multimap$}2
  {=>}{$\Rightarrow$}2
  {alpha}{$\alpha$}1,
  breaklines=true,
  morecomment=[l]{--},%
  morecomment=[s]{{-}{-}},%
  morestring=[b]',%
  morestring=[b]",%
  morestring=[s]{`}{`},%
}

\title{FreeST: Context-free Session Types in a Functional Language}
\author{
  Bernardo Almeida
  \and
  Andreia Mordido
  \and
  Vasco T. Vasconcelos
  \institute{LASIGE, Faculdade de Ciências, Universidade de Lisboa, Portugal}
}
\def\titlerunning{\freest}
\def\authorrunning{Almeida, Mordido \& Vasconcelos}

\begin{document}
\maketitle
\begin{abstract}
  \freest{} is an experimental concurrent programming language. Based
  on a core linear functional programming language, it features
  primitives to fork new threads, and for channel creation and
  communication. A powerful type system of context-free session types
  governs the interaction on channels.
  The compiler builds on a novel algorithm for deciding type
  equivalence of context-free session types.
  This abstract provides a gentle introduction to the language and
  discusses the validation process and runtime system.
\end{abstract}

%%% Local Variables:
%%% mode: latex
%%% TeX-master: "main"
%%% End:

\section{Session types deserve to be free}

Session types have been long subject to the shackles of tail
recursion~\cite{DBLP:conf/concur/Honda93,DBLP:conf/esop/HondaVK98}. Regular
session-type languages bear the evident advantage of providing simple
algorithms to check type equivalence and subtyping. Given two types, a
fixed-point construction algorithm such as the one introduced by Gay
and Hole builds, in polynomial time and space, a bisimulation relating
the two types, or decides that no such relation
exists~\cite{DBLP:journals/acta/GayH05}. The scenario darkens when one
decides to let go of tail recursion, for now the fixed-point
construction algorithm does not necessarily terminate.
This is one of the main reasons why session types have been confined
to ($\omega$-) regular languages for so many years.

The discipline of conventional (that is, regular) session types
provides guarantees not easily accessible to simpler languages such as
Concurrent ML, where channels are unidirectional and transport values
of a fixed type~\cite{DBLP:conf/mcmaster/Reppy93}. Session types, in
turn, provide for the description of richer protocols, epitomised by
the math client~\cite{DBLP:journals/acta/GayH05}, which can be
rendered in the SePi language~\cite{DBLP:conf/sefm/FrancoV13} as
follows:
\begin{lstlisting}[morekeywords=end]
MathClient = +{
  Plus: !Int.!Int.?Int.MathClient,
  Eq: !Int.!Int.?Bool.MathClient,
  Done: end
}
\end{lstlisting}

The type \lstinline|MathClient| describes the client side of a
protocol that introduces three choices: \lstinline|Plus|,
\lstinline|Eq|, and \lstinline|Done|. A client that chooses the
\lstinline|Plus| choice is supposed to send two integer values and to
receive a further integer (possibly representing the sum of the two
values sent), after which it goes back to the beginning. If the same
client chooses instead the \lstinline|Eq| option, it must subsequently
send two integer values and expect a boolean result (possibly
describing whether the two integers are equal), after which it must go
back to the beginning.  The \lstinline|Done| option terminates the
protocol, as described by type \lstinline[morekeywords=end]|end|.

The guarantees introduced by session type systems include the
adherence of code to protocols and the related absence of runtime
errors, including race
conditions~\cite{DBLP:conf/esop/HondaVK98}. Some systems further
guarantee progress (e.g.~\cite{DBLP:conf/concur/CairesP10}). All this,
under a rather expressive type language, that of (regular) session
types.

There is one further characteristic of session types that attest for
its flexibility: the ability to send channels on channels, often
called delegation. This feature provides for the transmission of
complex data on channels in a typeful manner. Suppose we want to
transmit a tree of integer values on a channel. The tree may be
described by data type \lstinline|Tree|:
\begin{lstlisting}
data Tree = Leaf | Node Int Tree Tree
\end{lstlisting}

One has to choose between a) using multiple channels for
sending the tree or b) using a single channel but incurring runtime
checks to check adherence to the protocol. In the former scenario,
trees are sent on channels of type
\begin{lstlisting}[morekeywords=end]
type TreeChannelC = +{Leaf:end, Node:!Int.!TreeChannelC.!TreeChannelC.end}
\end{lstlisting}
and we see that two new channels must be created and exchanged for
each \lstinline|Node| in a tree, so that $2n+1$ channels are needed to
transmit an $n$-\lstinline|Node| tree (equivalently, $n$ channels are
needed to transmit a tree with~$n$ \lstinline|Leaf| and
\lstinline|Node| components).
To see why, notice that type \lstinline|TreeChannelC| is not
recursive: a \lstinline|Leaf| is sent on the given channel, and so is
\lstinline|Node|, but two the two subtrees are sent on newly created
channels (witnessed by the parts \lstinline|!TreeChannelC| in the type).

In the latter case, tree parts are sent on a single channel, but not
necessarily in a ``tree form''. A suitable channel type in this case
is
\begin{lstlisting}[morekeywords=end]
type TreeParts = +{Leaf: TreeParts, Node: !Int.TreeParts, EOS: end}
\end{lstlisting}
where \lstinline|EOS| represents the end of transmission. In this case
tree \lstinline|Node 1 (Node 2 Leaf Leaf) Leaf| can be transmitted as
\lstinline|Node 1 Node 2 Leaf Leaf Leaf EOS|, when visited in a
depth-first manner. It should be easy to see that type
\lstinline|TreeParts| allows transmitting many different tree parts
that do not add up to a tree (such as,
\lstinline|Node 1 Leaf Leaf Node 2 Leaf EOS|), hence the necessary
runtime checks to look over unexpected parts on the channel.

In 2016, Thiemann and Vasconcelos introduced the concept of
context-free session types and proved that type equivalence remains
decidable~\cite{DBLP:conf/icfp/ThiemannV16}.
Context-free session types appear as a natural extension of
conventional (regular) session-types. Syntactically, the changes are
minor: rather than dot (\lstinline|.|), the prefix operator, we use
semi-colon (\lstinline|;|), a new binary operator on types. We also
take the chance to replace \lstinline [morekeywords=end]|end| by
\lstinline|Skip| to make it clear that it does not necessarily ends a
session type, but else merely introduces a mark that can sometimes be
omitted. In fact \lstinline|Skip| is the identity element of
sequential composition. The concurrent programming language we
present, \freest, is an implementation of the language proposed by
Thiemann and Vasconcelos.  \freest{} types enjoy the monoidal axioms,
\lstinline|Skip;S| $\equiv$ \lstinline|S;Skip| $\equiv$ \lstinline|S|,
associativity of sequential composition, \lstinline|S;(T;U)| $\equiv$
\lstinline|(S;T);U|, as well as the distributive property,
\lstinline|+{l:S, m:T};U| $\equiv$ \lstinline|+{l:S;U, m:T;U}| (and
similarly for external choice).
%for all types \lstinline|T|.
%
Using the \freest{} syntax, a channel that streams a tree can be
written as follows:
\begin{lstlisting}
type TreeChannel = +{Leaf: Skip, Node: !Int;TreeChannel;TreeChannel}
\end{lstlisting}

The language \freest, described in the remainder of the paper, 
provides for the best
of both worlds: stream the tree on a \emph{single} channel,
\emph{without} extraneous runtime checks.

% Related work

There are a few experimental programming languages based on session
types and there are many proposals for encoding session types in
mainstream programming languages. We cannot cover them all in
this short abstract; the interested reader is referred to a 2016
survey on behavioural types in programming
languages~\cite{DBLP:journals/ftpl/AnconaBB0CDGGGH16}.  Here, we briefly
discuss a few prototypical programming languages using session types.

SePi is a programming language based on the pi-calculus whose channels
are governed by regular session types refined with uninterpreted
predicates~\cite{DBLP:conf/sefm/FrancoV13}. The concrete syntax we
choose for \freest{} types is influenced by SePi.

SILL is a functional language with session typed concurrency, based on
the Curry-Howard interpretation of intuitionistic linear
logic~\cite{DBLP:conf/concur/CairesP10}, further extended with
recursive types and
processes~\cite{Toninho:phd,DBLP:conf/esop/ToninhoCP13}.
C1 is an imperative language~\cite{Pfenning:C1} developed along the
lines of SILL, featuring types that express
sharing~\cite{DBLP:journals/pacmpl/BalzerP17}.
All these languages use regular session types. Compared to \freest,
they present stronger properties (including progress), at the expense
of imposing a particular form of programming (derived from the
Curry-Howard interpretation) whereby each process uses zero or more
channels and provides exactly one. Unlike these languages, at the time
of this writing, \freest{} does not allow for multiple threads to
share a same channel.

Links is a functional programming language for the web, later extended
with session typing
primitives~\cite{DBLP:conf/fmco/CooperLWY06,DBLP:conf/esop/LindleyM15,Lindley.Morris_Lightweight.functional.session.types}. Very
much like SILL and C1, the base language is deadlock-free and
terminating. Links also includes recursion and shared channels, the
latter foregoing termination and deadlock-freedom.
The kinding system of \freest{} is quite similar to that of FST,
proposed by Lindley and
Morris~\cite{Lindley.Morris_Lightweight.functional.session.types}.

The only other implementation of context-free session types we are
aware of is that of Padovani~\cite{DBLP:conf/esop/Padovani17}, a
language that admits equivalence, subtyping and, of particular
interest, inference algorithms. It does however require a structural
alignment between the process code and the session types, enforced by
a \emph{resumption} process operator that explicitly breaks a type
$S;T$. The usage of resumptions requires additional annotations in
programs---something we managed to avoid in \freest---and,
furthermore, it does not solve the type equivalence problem since the
monoidal, associativity and distributivity properties typical of
context-free session types are not accounted for. Padovani proposes
the use of \emph{explicit coercions} to overcome this limitation, but
this requires greater efforts from programmers.  Additional effort by
programmers are always time consuming and error-prone, hence we have
embedded a full-fledged type equivalence checker into the compiler.

%%% Local Variables:
%%% mode: latex
%%% TeX-master: "main"
%%% End:

\section{\freest{} is for programming}
\label{sec:programming}
 
\freest{} is a basic implementation of the language introduced by
Thiemann and Vasconcelos~\cite{DBLP:conf/icfp/ThiemannV16}.
We have chosen the concrete syntax to be aligned with that of Haskell,
as much as possible. A \freest{} program is a collection of type
abbreviations, datatype and function (or value) declarations. Function
\lstinline|main| runs the program.

Our example serializes a tree object on a channel. The aim is to
transform a tree by interacting with a remote server. The client
process streams a tree on a (single) channel. In addition, for each
node sent, an integer is received. The server process reads a tree
from the other end of the channel and, for each node received, sends
back the sum of the integer values under (and including) that
node.
\newcommand{\leaf}{$\bullet$}
As an example, our program transforms the tree on the left into the one
on the right, where we use \leaf{} to abbreviate \lstinline|Leaf|.

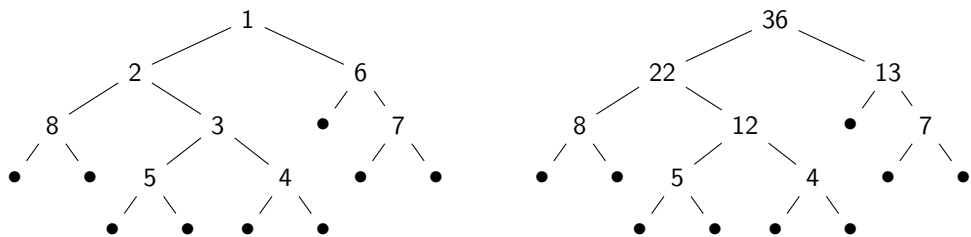
\begin{center}
  \begin{tikzpicture}[level 1/.style={sibling distance=3cm}, level
    2/.style={sibling distance=2.2cm}, level 3/.style={sibling
      distance=1.8cm}, level 4/.style={sibling distance=1cm}, level
    distance = 7mm]
    \node{\lstinline|1|} child { node {\lstinline|2|} child [level
      3/.append style={sibling distance=1cm}]{ node {\lstinline|8|}
        child { node {\leaf}} child { node {\leaf}}} child { node
        {\lstinline|3|} child { node {\lstinline|5|} child { node
            {\leaf}} child { node {\leaf}}} child { node
          {\lstinline|4|} child { node {\leaf}} child { node {\leaf}}
        }}} child[level 2/.append style={sibling distance=1cm}] { node
      {\lstinline|6|} child { node {\leaf}} child[level 3/.append
      style={sibling distance=1cm}] { node {\lstinline|7|} child {
          node {\leaf}} child { node {\leaf}}}} ;
  \end{tikzpicture}
  \hspace*{2em}
  \begin{tikzpicture}[level 1/.style={sibling distance=3cm}, level
    2/.style={sibling distance=2.2cm}, level 3/.style={sibling
      distance=1.8cm}, level 4/.style={sibling distance=1cm}, level
    distance = 7mm]
    \node{\lstinline|36|} child { node {\lstinline|22|} child [level
      3/.append style={sibling distance=1cm}]{ node {\lstinline|8|}
        child { node {\leaf}} child { node {\leaf}}} child { node
        {\lstinline|12|} child { node {\lstinline|5|} child { node
            {\leaf}} child { node {\leaf}}} child { node
          {\lstinline|4|} child { node {\leaf}} child { node {\leaf}}
        }}} child[level 2/.append style={sibling distance=1cm}] { node
      {\lstinline|13|} child { node {\leaf}} child[level 3/.append
      style={sibling distance=1cm}] { node {\lstinline|7|} child {
          node {\leaf}} child { node {\leaf}}}} ;
  \end{tikzpicture}
\end{center}

% The datatype for trees, \lstinline|Tree|, was introduced before.  
We need a channel capable of transmitting a tree. The type below
describes a variant of the \lstinline|TreeChannel| introduced in the
previous section that not only sends a tree, but, for each
\lstinline|Node|, also receives back an integer value.
\begin{lstlisting}
type TreeC = +{Leaf: Skip, Node: !Int;TreeC;TreeC;?Int}
\end{lstlisting}

The \lstinline|+| type constructor introduces an internal choice (the
process in possession of the channel end chooses) with two
alternatives, labelled \lstinline|Leaf| and \lstinline|Node|. These
two labels should not be confused with the constructors of datatype
\lstinline|Tree|. The \lstinline|Leaf| branch states that no further
interaction is possible on the channel (denoted by \lstinline|Skip|);
the \lstinline|Node| branch writes an integer value, followed by two
trees, and terminates by reading an integer value
(\lstinline|!Int;TreeC;TreeC;?Int|).
The type declaration introduces an \emph{abbreviation} to a recursive
type \lstinline|rec x. +{Leaf: Skip, Node: !Int;x;x;?Int}|.
This recursive type is not valid in conventional session type systems
given the non tail-recursive nature of the occurrences of
\lstinline|x|.

The writer process, \lstinline|transform|, writes a tree on a given
channel. It receives a tree of type \lstinline|Tree|  and a channel
of type \lstinline|TreeC;alpha|, for a type variable
\lstinline|alpha|. It returns a \lstinline|Tree| and the residual of
the input channel, of type \lstinline|alpha|. The type of
\lstinline|transform| is \emph{polymorphic}: different calls to the
function use different values for \lstinline|alpha|, as we show below.

\begin{lstlisting}
transform: forall alpha => Tree -> TreeC;alpha -> (Tree, alpha)
\end{lstlisting}

For each \lstinline|Node| in the input tree, function
\lstinline|transform| reads an integer from the channel and returns a
tree isomorphic to the input where the integer values in nodes are
read from the channel.
The function performs a \lstinline|case| analysis on the
\lstinline|Tree| constructor (either \lstinline|Leaf| or
\lstinline|Node|). In the former case, it selects the \lstinline|Leaf|
choice and returns a pair composed of the original tree and the
residual of the channel. In the latter case, the function selects the
\lstinline|Node| choice, then sends the integer value, followed by the
two subtrees (via recursive calls). Finally, it reads an integer
\lstinline|y| from the channel, assembles a tree with \lstinline|y| at
the root and returns this tree together with the residual of the
channel.
\begin{lstlisting}[numbers=left, xleftmargin=0.7cm]
transform tree c =
  case tree of
    Leaf ->
      (Leaf, select Leaf c)
    Node x l r ->
      let c   = select Node c in
      let c   = send x c in
      let l,c = transform[TreeC;?Int;alpha] l c in
      let r,c = transform[?Int;alpha] r c in
      let y,c = receive c in
      (Node y l r, c)
\end{lstlisting}

Notice that the language requires a continuous rebinding of channel
\lstinline|c| (lines 6--10), for its type changes at each interaction,
as in Gay and Vasconcelos~\cite{DBLP:journals/jfp/GayV10}.
At this point in time, \freest{} is not able to infer the appropriate
types that instantiate the polymorphic type variable \lstinline|alpha|
in the type of function \lstinline|transform|. We thus help the
compiler by supplying these types (\lstinline|TreeC;?Int;alpha| and
\lstinline|?Int;alpha|) in lines~8 and~9.

The reader process, \lstinline|treeSum|, reads a tree from a given
channel, writes back on the channel the sum of the elements in the
tree, and finally returns the sum. This process sees the channel from
the other end: rather than performing an internal choice
(\lstinline|+|), it performs an external choice (\lstinline|&|),
rather than writing (\lstinline|!|), it reads (\lstinline|?|), and
rather than reading, it writes. We abbreviate the thus obtained type,
\lstinline|rec x. &{Leaf: Skip, Node: ?Int;x;x;!Int}|, as
\lstinline|TreeS|. We say that \lstinline|TreeS| is \emph{dual} to
\lstinline|TreeC| and conversely. The signature of the reader process
is as follows.
\begin{lstlisting}
treeSum: forall alpha => TreeS;alpha -> (Int, alpha)
\end{lstlisting}

Rather than performing a case analysis on a \lstinline|Tree| as in the
writer process, function \lstinline|treeSum| matches the channel
against its two possible choices (\lstinline|Leaf| and
\lstinline|Node|). In the former case the function returns
\lstinline|0| (the sum of the integer values in an empty tree) and the
residual channel. In the latter, the function reads an integer from
the channel, then reads two subtrees (via recursive calls) and sends
on the channel the sum of the values in the subtree. It finally
returns this exact sum, together with the residual channel.
\label{lst:treeSum}
\begin{lstlisting}[numbers=left,firstnumber=12, xleftmargin=0.7cm]
treeSum c =
  match c with
    Leaf c ->
      (0, c)
    Node c ->
      let x, c = receive c in
      let l, c = treeSum[TreeS;!Int;alpha] c in
      let r, c = treeSum[!Int;alpha] c in
      let c    = send (x + l + r) c in
      (x + l + r, c)
\end{lstlisting}

Again notice the continuous rebinding of channel \lstinline|c| (lines
17--20) and the explicit types that instantiate variable
\lstinline|alpha| in the two recursive calls
(\lstinline|TreeS;!Int;alpha| and \lstinline|!Int;alpha|) in lines~18
and~19.

Function \lstinline|main| completes the program. It begins by creating
a new channel (line 24). The \lstinline|new| constructor takes a type
\lstinline|S| and returns a pair of channel ends of type
\lstinline|(S, T)|, where \lstinline|T| is a type dual to
\lstinline|S|. Then the \lstinline|main| function forks a new thread
to compute the \lstinline|treeSum| (line 25). In the main thread, it
transforms a given tree (\lstinline|aTree|). Function
\lstinline|treeSum| uses the \lstinline|r| end of the channel and
\lstinline|transform| uses \lstinline|w|, the other end. In these
calls, both functions are applied to type \lstinline|Skip| (lines
25--26). Analysing the signatures of the two functions (they both
return a channel of type \lstinline|alpha|), we see that the channel
ends \lstinline|r| and \lstinline|w| are both consumed to
\lstinline|Skip|. Type \lstinline|Skip| is unrestricted in nature (of
kind unrestricted) and hence its values can be safely discarded (cf.\
the two wilcards in the lets on lines 25--26). In addition to the
residual of channel end \lstinline|w|, function \lstinline|transform|
also returns a new tree \lstinline|t|, which becomes the result of the
\lstinline|main| function.
\begin{lstlisting}[numbers=left,firstnumber=22, xleftmargin=0.7cm]
main: Tree
main =
  let w,r = new TreeC in
  let _   = fork (treeSum[Skip] r) in
  let t,_ = transform[Skip] aTree w in
  t
\end{lstlisting}

When run, function \lstinline|main| prints on the console the textual
representation of the tree on the right of the above diagram if
\lstinline|aTree| denotes the tree on the left.

%%% Local Variables:
%%% mode: latex
%%% TeX-master: "main"
%%% End:

\section{Only valid programs may run}
\label{sec:valid}

This section introduces the main syntactic categories of the language:
kinds, types, expressions, and programs. It also discusses type
equivalence.

\paragraph{Kinding validates types}

\freest{} requires kinding; GV does. And the reason is on
polymorphism, not on context-free types.
\lstinline|!Int| is undoubtedly a session type, and so is
\lstinline|!Int;?Bool|. On the other hand, \lstinline|Int->Bool| and
\lstinline|(Int,Bool)| are certainly functional types. The GV language
allows a stratified grammar with separate syntactic categories for
functional types and for session types, even if mutually recursive.
In \freest{} this is not possible.  To see why, consider a type
variable \lstinline|alpha|. Is \lstinline|!Int;alpha| a session type?
What about \lstinline|alpha| itself? The former is a session type only
when \lstinline|alpha| represents a session type; the latter is a
functional type when \lstinline|alpha| denotes a functional type, and
is a session type otherwise. If \lstinline|alpha| does not denote a
session type, then \lstinline|!Int;alpha| is not a type, it is simply
a piece of useless syntax.

To accommodate polymorphism, types are classified into \emph{kinds}.
%
% Kinds classify types. 
Kinds are pairs composed of a \emph{prekind} and a
\emph{multiplicity}. Prekinds distinguish functional types,
\lstinline|T|, from session types, \lstinline|S|. Multiplicities
\begin{wrapfigure}{r}{0.12\textwidth}
  \begin{tikzpicture}[scale=.67]
    \node (TL) at (0,1) {\lstinline|TL|};
    \node (TU) at (-1,0) {\lstinline|TU|};
    \node (SL) at (1,0) {\lstinline|SL|};
    \node (SU) at (0,-1) {\lstinline|SU|};
    \draw (TL) -- (TU) -- (SU) -- (SL) -- (TL);
  \end{tikzpicture}
\end{wrapfigure}
control the number of times a value may be used in a
given context: exactly one---linear, \lstinline|L|---or zero or
more---unrestricted, \lstinline|U|. Both prekinds and multiplicities
come equipped with an ordering relation. Together they form the
lattice in the diagram on the right.
This is essentially the Lindley and Morris kinding system when
restricted to prekinds and
multiplicities~\cite{Lindley.Morris_Lightweight.functional.session.types}
or the Thiemann and Vasconcelos system where kinds for type schemes
are elided~\cite{DBLP:conf/icfp/ThiemannV16}.
Types of kind \lstinline|TU| can be used at kind
\lstinline|TL|. Similarly
types of kind \lstinline|SL| can be used at
kind \lstinline|TL|.  Finally, any type can be seen of kind
\lstinline|TL|, the kind sitting at top of the hierarchy, the kind
that carries the least information.

Equipped with a kinding system, \freest{} features as functional types
the following:
\begin{itemize}
\item Basic types, \lstinline|B : TU|, that is, \lstinline|Int|,
  \lstinline|Bool|, \lstinline|Char|, and \lstinline|()|,
\item Unrestricted and linear functions, \lstinline|T1 -> T2 : TU| and
  \lstinline|T1 -o T2 : TL|,
\item Pairs \lstinline|(T1, T2)| of kind
  depending on the kinds of
  \lstinline|T1| and \lstinline|T2|, and
\item Datatypes, \lstinline|[l1: T1, ..., ln: Tn]| of kind
  \lstinline|TU| or \lstinline|TL| depending on the kinds of the
  various types \lstinline|Ti|.
\end{itemize}

The session types are the following:
\begin{itemize}
\item The terminated type, \lstinline|Skip : SU|,
\item Sequential composition, \lstinline|S1;S2| of kind \lstinline|SU|
  or \lstinline|SL| depending on the kinds of \lstinline|S1| and
  \lstinline|S2|,
\item  Messages, \lstinline|!B, ?B|, both of kind \lstinline|SL|,
\item Choices, \lstinline|+{l1: S1, ..., ln:Sn}, &{l1: S1, ..., ln:Sn}|,
  both of kind \lstinline|SL|,
\item Recursive types, \lstinline|rec x.S|, bearing the kind of
  \lstinline|S|.
\end{itemize}

\paragraph{One function, many forms}

We are now in a better position to understand the type signature for
the functions in Section~\ref{sec:programming}.  Polymorphic variables
are introduced solely with the \lstinline|forall| construct. The
non-abbreviated type for function \lstinline|treeSum| spells out the
kind for the polymorphic type variable \lstinline|alpha|, as follows:
\begin{lstlisting}
forall alpha:SL => Tree -> TreeC;alpha -> (Tree,alpha)
\end{lstlisting}
The absence of an explicit kind for a polymorphic variable is
understood as \lstinline|SL|. In this particular case, one can replace
\lstinline|alpha| by any type with kind \lstinline|SL| or smaller,
including types \lstinline|TreeC;?Int;alpha : SL| (line 8 in the code
for \lstinline|TreeSum| in Section~\ref{sec:programming}),
\lstinline|?Int;alpha : SL| (line 9), and \lstinline|Skip : SU| (line
25 in the code for \lstinline|main|).
The call to function \lstinline|transform| on line
8 effectively calls the function at type
\begin{lstlisting}
Tree -> TreeC;TreeC;?Int;alpha -> (Tree, TreeC;?Int;alpha).
\end{lstlisting}
making it clear that channel \lstinline|c| is supposed to convey two
trees (the left and the right subtrees of a non-empty tree) before it
produces an integer value and continue as \lstinline|alpha|. The
function reads one tree from the channel and returns the unused part
of the channel, namely \lstinline|TreeC;?Int;alpha|.

Because we declared \lstinline|alpha| of kind \lstinline|SL|,
functional types cannot replace \lstinline|alpha|.  If one tries to
replace \lstinline|alpha| by \lstinline|Int->Bool|, one would get
\lstinline|Tree -> TreeC;(Int->Bool) -> (Tree,Int->Bool)|, which is
not a well formed type for the semicolon operator is defined on
session types only.
At the time of this writing datatypes are monomorphic.

\paragraph{Repeated behavior}

A characteristic central to session types is their ability to describe
unbounded behavior, usually captured by recursive types. Recursion is
certainly useful in types such as \lstinline|TreeC| in
Section~\ref{sec:programming}, meant do describe channels able to
transmit binary trees of different sizes. We have seen that name
\lstinline|TreeC| abbreviates type
\lstinline|rec x. +{Leaf: Skip, Node: !Int;x;x;?Int}|.  Type variable
\lstinline|x|, monomorphic, belongs to kind \lstinline|SU| so that
type
\lstinline|+{Leaf: Skip, Node: !Int;x;x;?Int}| may be deemed well
formed (and of kind \lstinline|SL|).
Recursive types must be \emph{contractive} or free from unguarded
recursion. The following types are not contractive:
\lstinline|rec x1. ... rec xn.x1| and
\lstinline|rec x1. ... rec xn.(x1;S)|, for \lstinline|n>1| and for any
session type \lstinline|S|.
Only session types can be recursive. Recursion in the functional part
of the type language is obtained via datatypes as usual, and the
fixed-point operator is built into function definition.

\paragraph{Expressions}

\freest{} blends expressions typical of functional languages and of
session types. In a way, it is not much different from GV.
The expressions inspired from functional languages include:
\begin{itemize}
\item Basic values: characters, integer and boolean values, and the
  unit value, \lstinline|()|,
\item Term variables (as opposed to type variables),
\item Lambda introduction, \lstinline|\x -o E| for linear abstractions
  and \lstinline|\x -> E| for unrestricted abstractions, and
  elimination, \lstinline|E1 E2|,
\item Pair introduction, \lstinline|(E1,E2)|, and elimination,
  \lstinline|let x, y = E1 in E2| (linear versions only, at the time
  of this writing),
\item Datatype elimination,
  \lstinline|case E of C1 x11...x1k -> E1, ..., Cn xn1...xnk -> En|, and
\item Conditional, \lstinline|if E1 then E2 else E3|, type
  application, \lstinline|x[T1,...Tn]|, and thread creation,
  \lstinline|fork E|.
\end{itemize}

The session-type related expressions are:
\begin{itemize}
\item Channel creation, \lstinline|new S|,
\item Message sending and receiving, \lstinline|send E| and
  \lstinline|receive E|, and
\item Branch selection, \lstinline|select C E|, and match,
  \lstinline|match E with C1 x -> E1,...,Cn x -> En|.
\end{itemize}

\paragraph{Programs}

Programs are composed of
\begin{itemize}
\item Function signatures and
  declarations, 
\item Datatype declarations, and
\item Type abbreviations.
\end{itemize}

Both functions and datatypes can be mutually recursive. Haskell code
is generated for programs that contain a function named
\lstinline|main| with a non-function type. The result of the
evaluation of this function is printed on the console.

%%% Local Variables:
%%% mode: latex
%%% TeX-master: "main"
%%% End

% \section{\freest{} requires a type equivalence checker}
% \label{sec:equivalence}

\paragraph{Checking type equivalence}

is the main challenge of the compiler. If, on the one hand, the
algorithm should be sound and complete, on the other hand it should
have a running time compatible with a compiler.  % However, the
% recursive structure underlying context-free session types often make
% bisimulations between two types infinite.

We have developed an algorithm to decide the equivalence of
context-free session types~\cite{typeEquivalence}.  It features three
distinct stages.
The \emph{first stage} builds a context-free grammar in Greibach
Normal Formal % ---in fact a simple grammar---
from a context-free session type in a way that bisimulation is
preserved.
The \emph{second stage} prunes the grammar by removing unreachable
symbols in unnormed sequences of non-terminal symbols. This stage
builds on the result of Christensen, H\"uttel, and 
Stirling~\cite{DBLP:journals/iandc/ChristensenHS95}.
The \emph{third stage} constructs an expansion tree, by alternating
between expansion and simplification steps. During an expansion step, 
all children nodes are expanded according to the transitions in the 
grammar. % Since we have simple grammars, no branching is expected at the
% expansion phase.
Throughout the simplification phase, the reflexive, congruence and
Basic Process Algebra (BPA) rules proposed by Janc\v ar and Moller
in~\cite{janvcar1999techniques} are applied to each node, yielding a
number of sibling nodes.  The simplification phase promotes branching
on the expansion tree.  The tree is traversed using breadth-first
search and the algorithm terminates as soon as it reaches an empty
node---case in which it decides positively---or it fails to expand a
node---case in which it decides negatively.  The finite representation
of bisimulations of BPA transition
graphs~\cite{caucal1986decidabilite,
  DBLP:journals/iandc/ChristensenHS95} is paramount for the soundness
and completeness of the algorithm, a result that we show to hold.

Although the branching nature of the expansion tree confers an
exponential complexity to the algorithm, we propose heuristics that
allow constructing the relation in a reasonable time. For this
purpose, we use a double-ended queue that allows prioritizing nodes
with potential to reach an empty node faster.  We have also speeded up
the computation of the expansion tree by iterating the simplification
phase until a fixed point is reached.  These optimizations led to an
improvement of more than $11,000,000\%$ on the runtime of the
algorithm, so that it can now be effectively incorporated in the
compiler for \freest.

%We have benchmarked the algorithm 
%on a test suite of carefully crafted pair of types comprising valid and invalid 
%equivalences, for a total os 138 tests. By running the tests on the base and
%optimized algorithms, we observe an improvement of more than 
%$11,000,000\%$ on runtime and memory allocated: it goes from 4445.38 
%seconds and 8,259,115 Mb memory allocated
%with the base algorithm to an average of 0.04
%seconds for the running time and 62 Mb of allocated memory with the 
%optimized algorithm.

%%% Local Variables:
%%% mode: latex
%%% TeX-master: "main"
%%% End:

\section{Let \freest{} run}
\label{sec:compiler}

\freest{} is written in Haskell and generates Haskell code that can be
compiled with a conventional GHC compiler.
The validation phase features a kinding system and a typing system
briefly described in the previous section. Only declarative versions
of these systems are described in Thiemann and
Vasconcelos~\cite{DBLP:conf/icfp/ThiemannV16}; we have developed the
corresponding algorithmic versions.

This section concentrates on the \emph{runtime system}, a surprisingly
compact system. We build on modules \lstinline|Control.Concurrent| and
\lstinline|Unsafe.Coerce|, and make particular use of the monadic
combinators below~\cite{DBLP:conf/afp/Wadler95}. The
\lstinline[language=Haskell]|do| notation is built on top of these
combinators.
\begin{lstlisting}[language=Haskell]
(>>) :: Monad m => m a -> m b -> m b
(>>=) :: Monad m => m a -> (a -> m b) -> m b
return :: Monad m => a -> m a
\end{lstlisting}

To fork a new thread, we use the forkIO primitive. The primitive
returns a threadId, which we discard and convert into the unit value.
\begin{lstlisting}[language=Haskell]
fork :: IO () -> IO ()
fork e = forkIO e >> return ()
\end{lstlisting}

Each channel is implemented with a pair of \lstinline|MVar|s first
introduced in Concurrent ML~\cite{DBLP:conf/mcmaster/Reppy93} and made
available in Haskell via module
\lstinline|Concurrent.Control|~\cite{DBLP:conf/popl/JonesGF96}. Each
channel end is itself a pair of crossed \lstinline|MVar|s.
\begin{lstlisting}
type ChannelEnd a b = (MVar a, MVar b)
\end{lstlisting}

The \lstinline|new| function creates such a pair and returns the two
channel ends in a pair.
\begin{lstlisting}[language=Haskell, literate={<-}{{$\leftarrow$}}1]
new :: IO (ChannelEnd a b, ChannelEnd b a)
new = do
  m1 <- newEmptyMVar
  m2 <- newEmptyMVar
  return ((m1, m2), (m2, m1))
\end{lstlisting}
% new = newEmptyMVar >>= \m1 -> newEmptyMVar >>= \m2 -> return ((m1, m2), (m2, m1)) 

To write on a channel end, we use the second \lstinline|MVar| in the
given pair. Because the type of values transmitted on a channel vary
over time, we use \lstinline|unsafeCoerce| to bypass the Haskell type
system, a technique common in standard implementations of session
types, including that of Pucella and
Tov~\cite{DBLP:conf/haskell/PucellaT08}. To write on the channel end
we use the \lstinline|putMVar| primitive. If the \lstinline|MVar| is
currently full, \lstinline|putMVar| waits until it becomes empty.  The
result of \lstinline|send| is the original channel end, which must be
rebound in programs, as discussed in the previous section.
\begin{lstlisting}[language=Haskell, literate={<-}{{$\leftarrow$}}1]
send :: b -> ChannelEnd a b -> IO (ChannelEnd a b)
send x (m1, m2) = putMVar m2 (unsafeCoerce x) >> return (m1, m2)
\end{lstlisting}

Finally, to read from a channel end, we use the first element of the
pair. The \lstinline|takeMVar| primitive returns the contents of the
\lstinline|MVar|. If the \lstinline|MVar| is empty,
\lstinline|takeMVar| waits until it is full. After a
\lstinline|takeMVar|, the \lstinline|MVar| is left empty. The result
of \lstinline|receive| is a pair composed of the value read from the
\lstinline|MVar| and the original channel (which, again, should be
rebound in the program).
\begin{lstlisting}[language=Haskell, literate={->}{{$\rightarrow$}}1]
receive :: ChannelEnd a b -> IO (a, ChannelEnd a b)
receive (m1, m2) = takeMVar m1 >>= \x -> return (unsafeCoerce x, (m1, m2))
\end{lstlisting}

Our implementation is \emph{not} completely aligned with the
operational semantics of \emph{synchronous} session types (and that
proposed by Thiemman and Vasconcelos in particular). We use
\lstinline|MVar|s to implement \emph{asynchronous} communication with
buffers of size one. The discrepancy can be witnessed with the standard
cross write-read on two channels. In the following program
\lstinline|w1|-\lstinline|r1| and \lstinline|w2|-\lstinline|r2| are
two channels.
\begin{lstlisting}
writer :: !Char -> !Bool -> Skip
writer w1 w2 =
  let _ = send 'c' w1 in
  send False w2
reader :: ?Char -> ?Bool -> Bool
reader r1 r2 =
  let x, r2 = receive r2 in
  let _, r1 = receive r1 in
  x
\end{lstlisting}
The program does not deadlock in our current implementation but does
so in a synchronous semantics. However, deadlock occurs in a simple
adaptation of the program with two consecutive writes two consecutive
reads on each channel (using a \lstinline|writer| of type
\lstinline|!Char;!Char -> !Bool;!Bool -> Skip| and dually for
\lstinline|reader|).

% The semantics of Haskell \lstinline|MVar|s are completely aligned with
% that of the communication primitives of (synchronous) session types. O
% \lstinline|MVar|s implement a one-place buffer providing for the
% rendez-vous (the first to arrive at the read/write point waits)
% mechanism underlying session communication. Given the linear nature of
% \freest{} channels, no actual queues are required.

\freest{} is a standalone language whose types need not be in the
runtime system of Haskell, since type checking is performed by the
\freest{} front-end.
Coercions (that is, calls to the \lstinline|unsafeCoerce| function)
are then inserted so as to avoid typing errors when compiling the
target Haskell code. The occurrences of \lstinline|unsafeCoerce| are
limited to \emph{two} places: when reading and when writing from
\lstinline|MVar|s, in runtime functions \lstinline|receive| and
\lstinline|send|, respectively.

% Call-by-value and call-by-name.
The original proposal of context-free session types feature a
call-by-value semantics. In order to align \freest{} with this
semantics, we use Haskell's \emph{bang patterns}. For example, the
\lstinline|transform| function is compiled into
\lstinline|transform !tree !c = ...|, forcing the evaluation of both
parameters, prior to the execution of the function body, thus
effectively implementing a call-by-value strategy.

%%% Local Variables:
%%% mode: latex
%%% TeX-master: "main"
%%% End:

\section{The bright future of \freest{}}
\label{sec:conclusion}

We have developed a basic compiler for \freest, a concurrent
functional programming with context-free session types, based on the
ideas of Thiemann and Vasconcelos~\cite{DBLP:conf/icfp/ThiemannV16}.
There are many possible extensions to the language. We discuss a
few.
Support for linear pairs and linear datatypes, as well for polymorphic
datatypes, should not be difficult to incorporate.
Because \freest{} compiles to Haskell, a better interoperability is
called for. We plan to add primitive support for lists, and for some
functions in Haskell's prelude, namely rank-1 functions that we will
have to annotate with \freest{} types.
We have chosen a buffered semantics with buffers of size one, for ease
of implementation, but we plan to experiment with buffered channels of
arbitrary size by simply replacing the runtime system.
The original proposal of context-free sessions is based on a
call-by-value operational semantics and we kept that strategy in
\freest. We however plan to experiment with call-by-need, taking
advantage of the fact that \freest{} compiles to Haskell.
Shared channels allow for multiple readers and multiple writers,
thus introducing (benign) races. There are several proposals in the
literature~\cite{DBLP:journals/pacmpl/BalzerP17,
  DBLP:conf/sefm/FrancoV13,Lindley.Morris_Lightweight.functional.session.types,DBLP:journals/iandc/Vasconcelos12}
on which we may base this extension.
The \lstinline|dualof| type operator is present in the SePi
language~\cite{DBLP:conf/sefm/FrancoV13}; we plan its incorporation in
\freest{} language.

We also plan to extend the expressivity of \freest{} by allowing
messages to convey arbitrary types, as opposed to basic types only.
% session types to be used to send or receive other session types.  For
% this purpose, we intend to enable message operators to be applied not
% only to basic types but to any functional or session type. 
In this wider scope, the type equivalence algorithm for context-free
session types must be intertwined with the type equivalence algorithm
for functional types, for now the labels of the labeled-transition
system are types themselves.

Last but not least, we plan to incorporate type inference on type
applications in order to allow the automatic identification of the
unifier matching a polymorphic type against a given type. This
unification process should recognize the unifiers of two types up to
type equivalence. However, dealing with type inference on type
applications with recursive types might be challenging, as observed by
Hosoya and Pierce~\cite{DBLP:journals/toplas/PierceT00}.
%in~\cite{hosoya1999good}.

%%% Local Variables:
%%% mode: latex
%%% TeX-master: "main"
%%% End:

\paragraph{Acknowledgements} The authors would like to thank the
anonymous reviewers for their extremely relevant comments and
pointers.  This work was supported by FCT through project Confident
(PTDC/EEI-CTP/ 4503/2014), by the LASIGE Research Unit
(UID/CEC/00408/2019) and by Cost Action CA15123 EUTypes, supported by
COST (European Cooperation in Science and Technology).

\bibliography{biblio}
\bibliographystyle{eptcs}

\end{document}